\documentclass[showpacs,10pt,twocolumn,prb]{revtex4-1}

\usepackage{amsmath}
\usepackage{amssymb}
\usepackage{graphicx}
\usepackage{amssymb}
\usepackage{graphics}
\usepackage{epsfig}
\usepackage{CJK}
\usepackage{color}

\begin{document}

\title{Two-dimensional Dirac fermions in YbMnBi$_{2}$ antiferromagnet}
\author{Aifeng Wang,$^{1}$ I. Zaliznyak,$^{1}$ Weijun Ren,$^{1,2}$ Lijun Wu,$^{1}$ D. Graf,$^{3}$ V. O. Garlea,$^{4}$ J. B. Warren,$^{5}$, E. Bozin,$^{1}$ Yimei Zhu$^{1}$ and C. Petrovic$^{1}$}
\affiliation{$^{1}$Condensed Matter Physics and Materials Science Department, Brookhaven National Laboratory, Upton, New York 11973, USA\\
$^{2}$Shenyang National Laboratory for Materials Science, Institute of Metal Research, Chinese Academy of Sciences, Shenyang 110016, China\\
$^{3}$National High Magnetic Field Laboratory, Florida State University, Tallahassee, Florida 32306-4005, USA\\
$^{4}$Quantum Condensed Matter Division, Oak Ridge National Laboratory, Oak Ridge, Tennessee 37831, USA\\
$^{5}$Instrument Division, Brookhaven National Laboratory, Upton, New York 11973, USA}

\date{\today}

\begin{abstract}
We report two-dimensional quantum transport and Dirac fermions in YbMnBi$_{2}$ single crystals. YbMnBi$_{2}$ is a layered material with anisotropic conductivity and magnetic order below 290 K. Magnetotransport properties, nonzero Berry phase and small cyclotron mass indicate the presence of quasi two dimensional Dirac fermions. Quantum oscillations in Hall resistivity suggest the presence of both electron and hole parts of the Fermi surface whereas the Berry phase suggests spin-orbit coupling.
\end{abstract}
\pacs{72.20.My, 72.80.Jc, 75.47.Np}
\maketitle

\section{INTRODUCTION}

The energy disperson of carriers in Dirac materials can be approximated by the relativistic Dirac equation.\cite{Dirac} By now it has been established that Dirac states can be found in wide range of materials such as iron-based or copper oxide superconductors, graphene, and topological insulators.\cite{Vafek,PRL 104 137001,Science 288 468,NovoselovK,NM 6 183,RMP 81 109,RMP 82 3045,RMP 83 1057} In the quantum limit all carriers are condensed to the lowest Landau level (LL).\cite{Fundamentals} This is easily realized in laboratory magnetic fields for Dirac fermions since the distance between the lowest and first LL of Dirac fermions is large, in contrast to the conventional electron gas with parabolic energy dispersion. In such a case the components of the resistivity tensor $\rho_{xx}$ and $\rho_{xy}$ are linear in magnetic field,\cite{PRB 58 2788} quantum Hall effect, non-trivial Berry phase and large unsaturated linear magnetoresistance (MR) are observed.\cite{NovoselovK,Nature 438 201,Science 324 924,PRL 106 217004}

\begin{figure}
\centerline{\includegraphics[scale=0.40]{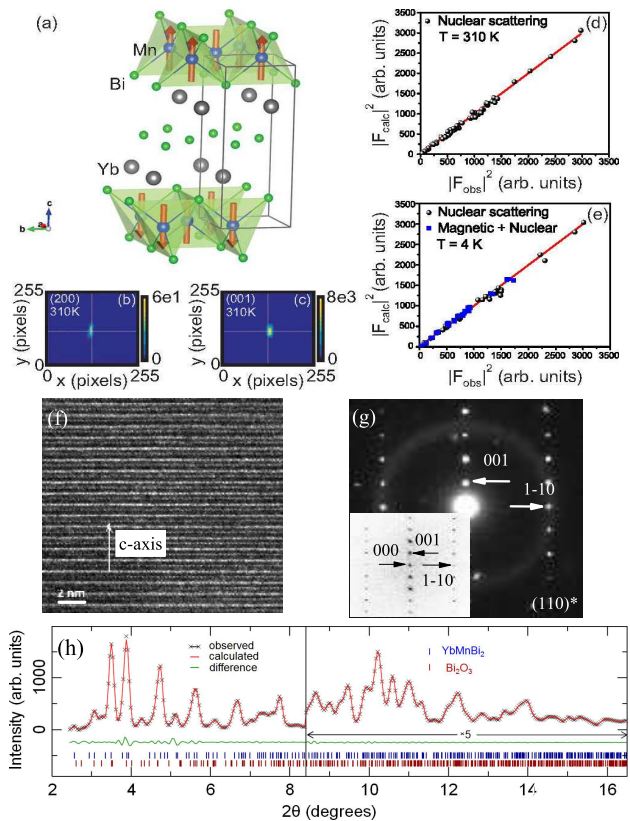}} \vspace*{-0.3cm}
\caption{(Color online). (a) Crystal and magnetic structure of YbMnBi$_{2}$. Arrows show the ordered Mn magnetic moments, $\mu_{Mn} = 4.3(1)\mu_B$, refined at 4 K. The moments point along the $c$ axis, the magnetic space group describing the AFM order is $P4'/n'm'm : (3/4,1/4,0 | 0,0,mz) (1/4,3/4,0 | 0,0,-mz)$. (b), (c) Neutron intensity patterns of (200) and (001) nuclear structural Bragg peaks, respectively, of a YbMnBi$_2$ single crystal on the two-dimensional position sensitive detector on HB3a diffractometer, indicating perfect crystalline structure. (d), (e) Summary of the structural refinement of YbMnBi$_2$ in the magnetically disordered phase at 310 K, and at 4 K, where it is antiferromagnetically ordered. (f) High resolution transmission electron microscopy (HRTEM) image and corresponding electron diffraction pattern (EDP) viewed along [110] direction (g). The inset in (g) is the FFT from the image shown in (f). (h) Structural refinement of powder diffraction data. Ticks mark reflections, top row refers to the main phase, bottom row refers to Bi$_{2}$O$_{3}$ due to sample preparation.}
\label{magnetism}
\end{figure}

Alkaline earth ternary AMnBi$_{2}$ crystals (A = alkaline earth such as Sr or Ca) have also been shown to host quasi-2D Dirac states similar to graphene and topological insulators.\cite{KefengSr,KefengCa,ParkSr,JiaLL,PRB 87 245104,sr 4 5385} The local arrangement of alkaline earth atoms and spin-orbit coupling are rather important for the characteristics of Dirac cone states. In SrMnBi$_{2}$ the degeneracy along the band crossing line is lifted except at the place of anisotropic Dirac cone. In contrast, the energy eigenvalue difference due to perturbation potential created by staggered alkaline earth atoms results in a zero-energy line in momentum space in CaMnBi$_{2}$.

Rare earth compounds RMnBi$_{2}$ (R=Eu,Yb) are isostructural to SrMnBi$_{2}$ and CaMnBi$_{2}$, respectively.\cite{Sales,CavaRJ} Whereas antiferromagnetic order of Eu atoms in EuMnBi$_{2}$ influences Dirac states below Neel temperature $T_{N}$ = 22 K and consenquently magnetotransport, Dirac semimetal YbMnBi$_{2}$ features time-reversal symmetry breaking Weyl semimetallic state where the Dirac node separates into two Weyl nodes.\cite{Balents,CavaRJ,BurkovAA,WanX} Here we report the quasi-two-dimensional quantum transport in YbMnBi$_{2}$. The nonzero Berry phase, small cyclotron mass and large mobility confirm the existence of Dirac fermions in Bi square nets. The quasi-2D in plane magnetoresistance (MR) shows a crossover from parabolic-in-field semiclassical MR to high-field linear-in-field dependence. The temperature dependence of crossover field $B^{*}$ is quadratic, as expected for Dirac fermions.

\section{EXPERIMENTAL DETAILS}

YbMnBi$_2$ single crystals were grown from excess Bi. Yb, Mn, and Bi were mixed together according to the ratio Yb: Mn: Bi = 1: 1: 10. Then, the mixture was placed into an alumina crucible, sealed in a quartz tube, heated slowly to 900 $^{\circ}$C, kept at 900 $^{\circ}$C for 2 h and cooled to 400 $^{\circ}$C, where the excess Bi flux was decanted. Shiny single crystals with typical size 3 $\times$ 3 $\times$ 1 mm$^3$ can be obtained. The single crystals free of residual flux droplets can be obtained by cutting the six faces of the cuboid. The element analysis was performed using an energy-dispersive x-ray spectroscopy (EDX) in a JEOL LSM-6500 scanning electron microscope. Single crystal neutron diffraction measurements were performed using the HB3A four-circle diffractometer at the High Flux Isotope Reactor at Oak Ridge National Laboratory. A crystal specimen of approximately 2 x 2 x 1 mm$^{3}$ was loaded in a closed-cycle-refrigerator whose temperature was controlled in the range (4 - 350) K. For the measurements we used a monochromatic beam with the wavelength 1.551 ${\AA}$ selected by a multilayer [110]-wafer silicon monochromator, and the scattered intensity was measured using an Anger-camera type detector. The neutron diffraction data were analyzed by the using the FullProf Suite package. Transmission-electron-microscopy (TEM) sample was prepared by crushing the single crystal sample, and then dropping to Lacey carbon grid. X-ray powder diffraction data were collected at 28-ID-C beamline of NSLS-II at Brookhaven National Laboratory by utilizing monochromatic beam ($\lambda$ = 0.01858 nm) and Perkin-Elmer image plate detector from pulverized sample placed in cylindrical polyimide capilary. Data integration to 2$\theta$ was carried out using Fit2D, while structural refinement of P4/nmm model used GSAS operated under EXPGUI platform.\cite{Hammersley,Larson,Toby} High-resolution TEM imaging and electron diffraction were performed using the double aberration-corrected JEOL-ARM200CF microscope with a cold-field emission gun and operated at 200 kV. Magnetotransport measurements up to 9 T were conducted in a Quantum Design PPMS-9 with a conventional four wire method on cleaved and polished single crystals. Polishing is necessary in order to remove residual bismuth droplets from the surface of as-grown single crystals. Magnetotransport at high magnetic field up to 35 T was conducted at National High Magnetic Field Laboratory (NHMFL) in Tallahassee.

\section{CRYSTAL AND MAGNETIC STRUCTURE}

The crystal and magnetic structure of YbMnBi$_{2}$ determined by neutron diffraction and high resolution TEM (HRTEM) is presented in Figure 1. The nuclear lattice structure was determined from measurements at $T = 310$~K, where magnetic order is absent. The magnetic structure was determined at $T = 4$~K, and the refined saturated magnetic moment at this temperature is $4.3(1)\mu_B/$Mn. Refinements were carried out using data sets of 82 reflections, the resulting structural parameters are listed in Table \ref{Table1-ND}. R-factors $\sim 5\%$ were obtained for both temperatures. No indication of structural transformation between 300K and 4K has been detected, and no orthorhombic or monoclinic distortions were observed within the HB3a wave vector resolution. The data  is fit equally well in both tetragonal and orthorhombic symmetry. The magnetic space group describing the antiferromagnetic (AFM) order at 4~K [Fig. 1 (a)] is $P4'/n'm'm : (3/4,1/4,0 | 0,0,mz) (1/4,3/4,0 | 0,0,-mz)$. Note that this symmetry does not allow canting of Mn magnetic moments away from the $c-$axis, such as inferred in Ref. \onlinecite{CavaRJ}. In fact, none of the maximal subgroups of $P4/nmm1'$ allows such canting, which requires a lower symmetry group, e.g. $Pm'n'21$, and which would mean that either the AFM transition is not of the second order, or that the lattice symmetry is not tetragonal, for which our measurements provide no indications. The SEM obtained atomic ratio of Yb: Mn: Bi is 26: 26: 48, consistent with the composition YbMnBi$_2$. Both HRTEM electron diffraction pattern and fast Fourier transform can be well indexed as (110)* zone of YbMnBi$_{2}$ structure. Powder diffraction data are well explained by \textit{P4/nmm} model of YbMnBi$_{2}$ [a=4.488(2) ${\AA}$, c=10.826(2) ${\AA}$), Fig. 1(h). In addition to the main phase, about 10\% by weight of Bi$_{2}$O$_{3}$ phase was also observed due to secondary oxidation of unreacted Bi metal on crystal surface during pulverization of the single crystal specimen.

\begin{table}[!ht]\footnotesize
\vspace{-0.25in}
\caption{The crystal and magnetic structure parameters of YbMnBi$_{2}$ determined by neutron diffraction. The refinement was carried out using 82 Bragg reflections measured on the HB3A diffractometer using monochromatic neutron beam with wavelength $\lambda \approx 1.55 \AA$. Each reflection was measured by performing the sample rotation (omega) scan to extract the integrated intensity. }\footnotesize%
\label{Table1-ND}
 \vspace{0.05in}
 \scalebox{1}{
\begin{tabular}[c]{lccccc}
 \hline\hline
 \multicolumn{6}{c}{$T = 310$~K. } \\
 \hline
 \multicolumn{6}{c}{Symmetry group: $P 4/n m m$. Bragg R-factor:  5.25.} \\
 \multicolumn{6}{c}{$a = b = 4.48(1)$, $c = 10.80(2)$. Magnetic moment $0\mu_B/$Mn. } \\
 \hline
Atom & x & y & z & $B_{iso}$ & Site Multiplicity \\
Yb1 & 0.25 & 0.25 & 0.73174(19) & 1.051(148) & 2 \\
Mn1 & 0.75 & 0.25 & 0.0 & 1.255(255) & 2 \\
Bi1 & 0.75 & 0.25 & 0.5 & 0.917(159) & 2 \\
Bi2 & 0.25 & 0.25 & 0.16700(29) & 1.027(158) & 2 \\
 \hline
 \multicolumn{6}{c}{$T = 4$~K. } \\
 \hline
 \multicolumn{6}{c}{Symmetry group: $P 4/n m m$. Bragg R-factor:  4.83.} \\
 \multicolumn{6}{c}{$a = b = 4.46(1)$, $c = 10.73(2)$. Magnetic moment $4.3(1)\mu_B/$Mn. } \\
 \hline
Atom & x & y & z & $B_{iso}$ & Site Multiplicity \\
Yb1 & 0.25 & 0.25 & 0.73143(21) & 0.379(173) & 2 \\
Mn1 & 0.75 & 0.25 & 0.0 & 0.707(272) & 2 \\
Bi1 & 0.75 & 0.25 & 0.5 & 0.188(194) & 2 \\
Bi2 & 0.25 & 0.25 & 0.16567(32) & 0.256(187) & 2 \\
 \hline\hline
\end{tabular}
}
\end{table}

\section{RESULTS AND DISCUSSION}

Figure 2(a) shows the temperature dependence of the in-plane ($\rho_{ab}$) and out-of-plane ($\rho_c$) resistivity at 0 and 9 T for  YbMnBi$_2$ single crystal. The in-plane resistivity becomes flat below 8 K, extrapolating to a residual resistivity $\rho_0$(0 T) $\approx$ 4.77 $\mu\Omega$ cm. The residual resistivity ratio (RRR) $\rho$(300 K)/$\rho_0$ is about 20. The MR ratio MR = [$\rho_{ab}(B)$-$\rho_{ab}(0)$] is 234\% at 2 K in a 9 T field. The MR is gradually suppressed with temperature increase. The resistivity  $\rho_c (T)$/$\rho_{ab} (T)$ is highly anisotropic. The hump below 300 K in $\rho_c (T)$, could indicate a crossover from high-T incoherent to low-T coherent conduction.\cite{Gutman1,Gutman2} As shown in inset in Fig. 2(a), $\rho_{ab}$ is quadratic-in-temperature below about 5 K: $\rho (T)$ = $\rho_0$ + $AT^2$ with A = 5.74 n$\Omega$ cm K$^{-2}$. The parameter A is inversely proportional to the Fermi temperature, and is only one third of that of SrMnBi$_2$.\cite{ParkSr} This indicates that the effective mass in YbMnBi$_2$ is rather small.

Specific-heat measurement on YbMnBi$_2$ is shown in Fig. 2(b). A peak is clearly observed at around 285 K, which could be attributed to the magnetic transition.\cite{GuoYF} The fitting of the low temperature data using $C_p$ = $\gamma_nT$ + $\beta$$T^3$ + $\eta$$T^5$ gives $\gamma_n$ = 2.16 mJ mol$^{-1}$ K$^{-2}$, $\beta$ = 2.36 mJ mol$^{-1}$ K$^{-4}$, and $\eta$ = 0.00695 mJ mol$^{-1}$ K$^{-6}$. Thus, Debye temperature of 149 K can be obtained.

\begin{figure}[ht]
\centering
\includegraphics[width=0.475\textwidth]{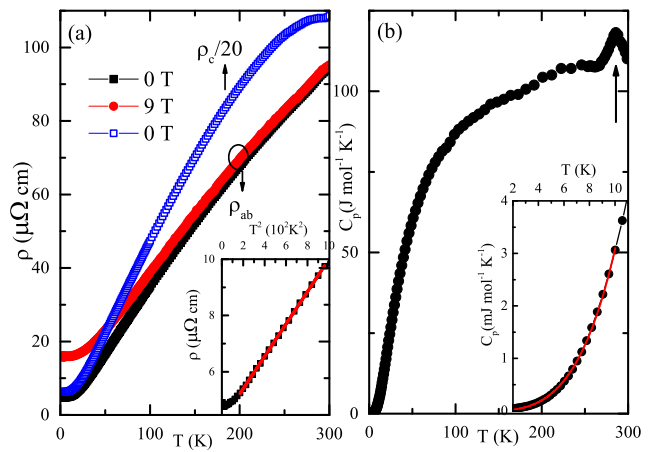}
\caption{(color online). (a) Temperature dependence of in-plane resistivity ($\rho_{ab}$) for YbMnBi$_2$ single crystals at 0 and 9 T, and out-of-plane resistivity ($\rho_{c}$) at 0 T. Inset shows the quadratic $T$ dependence at low temperature of $\rho_{ab}$ at 0 T. (b) Temperature dependence of the specific heat of YbMnBi$_2$, inset shows the fitting of the low-temperature part.}
\label{fig1}
\end{figure}

\begin{figure}[ht]
\centering
\includegraphics[width=0.5\textwidth]{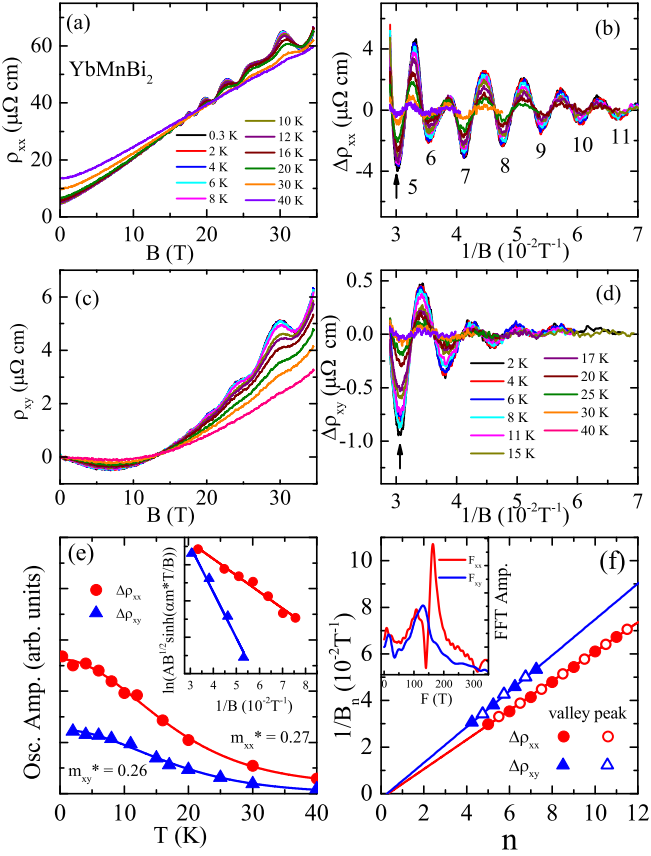}
\caption{(color online).   Longitudinal and transverse in-plane resistivity versus field at different temperatures (a-d). SdH oscillatory components $\Delta\rho_{xx,xy}$ are obtained after the background subtraction. (e) Temperature dependence of the oscillating amplitude of longitudinal (red circle) and transverse (blue triangle) at 1/B = 0.33 T$^{-1}$ and 0.34 T$^{-1}$, respectively. The solid line is the fitting curve. inset: Dingle plot for $\Delta\rho_{xx}$ and  $\Delta\rho_{xy}$ (f) LL index plots 1/B$_n$ versus $n$. The inset shows the Fourier transform spectrum of  $\Delta\rho_{xx}$ and  $\Delta\rho_{xy}$} \label{fig2}
\end{figure}

\begin{figure}[ht]
\centering
\includegraphics[width=0.45\textwidth]{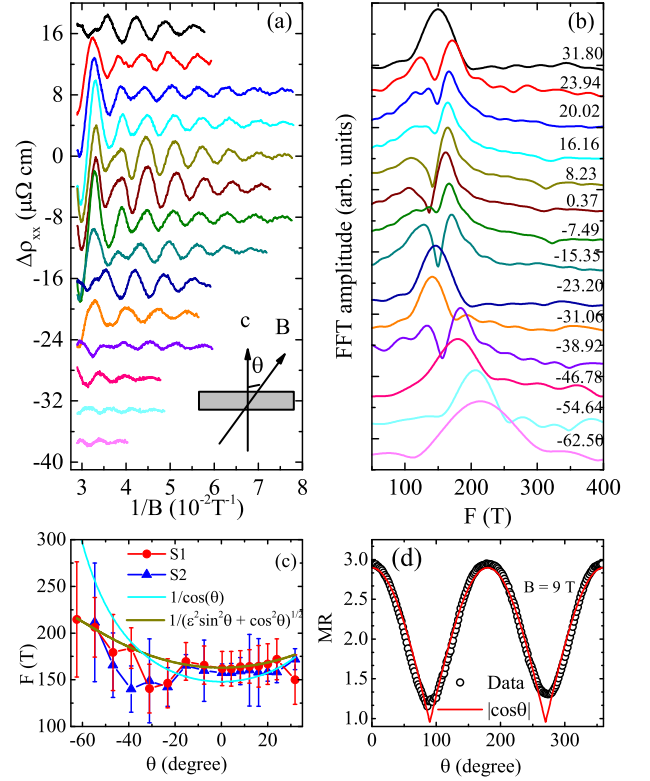}
\caption{(color online). (a)  The SdH oscillatory component as a function of 1/B at various angle, the cartoon shows the geometry of the measurement setup. (b) FFT spectra of the SdH oscillations in Fig. 4(a), (a) and (b) use the same legend. (c) Angular dependence of the oscillation frequency, Sample 1 (S1) is the sample we used in this paper, Sample 2 is the sample from same batch and measured on the same probe. Cyan line is the fitting using 2D model with $F(0)$/cos($\theta$), dark yellow line is the ellipse model $F(0)$/($\varepsilon^2$sin$^2\theta$ + cos$^2\theta$)$^{1/2}$ with an anisotropic factor $\varepsilon$ = -0.67. The error bar is the peak width at half height of the FFT peaks (d) Magnetoresistance as a function of the titled angle from applied field. The red line is the fitting using 2D model.} \label{fig3}
\end{figure}

Clear Shubnikov-de Haas (SdH) oscillations in both  longitudinal $\rho_{xx}$ and transverse $\rho_{xy}$ resistivity are observed up to 35 Tesla [Fig.3(a-d)]. From the fast Fourier transform (FFT) background-subtracted oscillating components $\Delta$$\rho_{xx}$ and $\Delta$$\rho_{xy}$ feature a single ($F$ = 130 T)and double frequency (164 T and 108 T), respectively, at 2 K [Fig. 3(b,d)]. According to the classical expression for the Hall coefficient when both electron- and hole-type carriers are present,\cite{Smith}
\[\begin{array}{l}
\frac{{{\rho _{xy}}}}{{{\mu _0}H}} = \mathop R\nolimits_H {\rm{ }}
 = \frac{1}{e}\frac{{(\mu _h^2{n_h} - \mu _e^2{n_e}) + {{({\mu _h}{\mu _e})}^2}{{({\mu _0}H)}^2}({n_h} - {n_e})}}{{{{({\mu _e}{n_h} + {\mu _h}{n_e})}^2} + {{({\mu _h}{\mu _e})}^2}{{({\mu _0}H)}^2}{{({n_h} - {n_e})}^2}}}
\end{array}\]

In the weak-field limit, the equation can be simplified as $R_H$ = $e^{-1}$$(\mu _h^2{n_h}-\mu _e^2{n_e})$/$(\mu _h{n_h}+\mu _e{n_e})^2$, while $R_H$ = 1/$(n_h-n_e)e$ in the high field limit. The positive slope of $\rho_{xy}(H)$ at high field gives $n_h > n_e$, and the negative slope of $\rho_{xy}(H)$ at weak field indicates  $(\mu _h^2{n_h}-\mu _e^2{n_e}) < 0$. Therefore, $\mu_e > \mu_h$ at low field, and the nonlinear behavior is the result of the carrier density and mobility competition between hole- and electron-type carriers. Fitting $\rho_{xy}(2 K)$ at high field using $R_H$ = 1/$(n_h-n_e)e$ yields $n_h - n_e$ = 2.09 $\times 10^{21}$ cm$^{-3}$. The mobility $\mu_{tr}$ = $R_H/\rho_{xx}$ is estimated to be 6.27  $\times 10^{6}$ cm$^2$ V$^{-1}$ s$^{-1}$ at 2 K, comparable to that in NbP and Cd$_3$As$_2.$\cite{NbP,LiangT}

The main frequency of longitudinal oscillation (164 T) is between the frequencies of SrMnBi$_2$ and CaMnBi$_2$. The difference in FFT frequencies between $\Delta$$\rho_{xx}$ and $\Delta$$\rho_{xy}$ is likely due to the different Fermi surface contribution in Hall resistivity channel.  From the Onsager relation $F$ = ($\Phi_0$/2$\pi^2$)$A_F$, where $\Phi_0$ is the flux quantum and $A_F$ is the orthogonal cross-sectional area of the Fermi surface, we estimate $A_F$ = 1.56 nm$^{-2}$.  This is rather small, similar to that in SrMnBi$_2$ (1.45 nm$^{-2}$) and is only a few \% of the total area of Brillouin zone in (001) plane.\cite{ParkSr,KefengCa} We can also approximate $k_F$$\approx$ 0.705 nm$^{-1}$, assuming the circular cross-section $A_F$ = $\pi$$k_F^2$. Angle-resolved photoemission spectroscopy (ARPES) and theory calculations reveal three small Fermi surface parts in SrMnBi$_2$:  needle-shaped FSs on the $\Gamma$ - $M$ line and near the $X$ point, and flower-shaped hole pockets near the $Z$ point.\cite{ParkSr,CavaRJ} It has been pointed out that the main frequency corresponds to the needle-shaped Fermi surfaces on the $\Gamma$ - $M$ line\, where a Dirac dispersion is expected.\cite{ParkSr,JoYJ} The flower-shaped pockets near the $Z$ point are larger than the FS on the $\Gamma$ - $M$ line. Assuming similar Fermi surface properties between YbMnBi$_{2}$ and SrMnBi$_{2}$, the smaller frequency of 108 T might be attributed to the small FS near the $X$ point, consistent with the theory calculation, which found small electron-like pockets near the X-points.\cite{CavaRJ}

Weyl points and the Fermi Arc connecting these points have been directly observed by ARPES in YbMnBi$_2$.\cite{CavaRJ} It is reported that the Fermi arcs, which participate in unusual closed magnetic orbits by traversing the bulk of the sample to connect opposite surfaces, can be detected by quantum oscillation.\cite{NC 5 5161} In our experiment we do not observe the frequency associated with the Fermi arc, possibly due to macroscopic thickness of our sample that exceeds the mean-free path.\cite{NC 5 5161, Moll}

The cyclotron masses and quantum life time of carriers can be extracted from the temperature and field dependence of oscillation amplitude using the Lifshitz-Kosevich formula.\cite{Shoeneberg}

\[\frac{{\Delta {\rho _{xx}}(T,B)}}{{{\rho _{xx}}(0)}} = {e^{ - \alpha m^*{T_D}/B}}\frac{{\alpha m^*T/B}}{{\sinh (\alpha m^*{T_D}/B)}}\]

Where $\alpha$ = 2$\pi$$^2$$k_{\rm B}$/e$\hbar$ $\approx$ 14.69 T/K, $m^{*}$ = $m$/$m_{e}$ is the cyclotron mass ratio ($m_{e}$ is the mass of free electron), and $T_{D} = \frac{\hbar}{2\pi{k_B}\tau_Q}$, with $\tau_Q$ the quantum lifetime. By fitting the thermal damping of the oscillation peak indicated by the arrow in Fig. 3(b) and 3(d), we can extract the cyclotron mass $m^{*}$ $\approx$ 0.27 and 0.26 for $\Delta$$\rho_{xx}$ and $\Delta$$\rho_{xy}$, similar to that in SrMnBi$_2$ and CaMnBi$_2$.\cite{ParkSr,KefengSr} Then, a very large Fermi velocity $\nu_F$ = $\hbar{k_F}/m^*$ = 3.01 $\times 10^{5}$ m/s can be obtained.  As shown in Fig. 3(e), $T_{D,xx}$ = 11.4 K and $T_{D,xy}$ = 31.81 K can be obtained by fitting the field dependence of the oscillation amplitude, and $T_{D,xx}$ is larger than that in BaMnBi$_2$.\cite{LiLJ} As a result, $\tau_{xx}$ = 1.05 $\times 10^{-13}$ s and $\tau_{xy}$ = 3.7 $\times 10^{-14}$ s are obtained. $\tau_{xx}$/$\tau_{xy}$ = 2.8, indicating the anisotropy in longitudinal and transverse scatting time.  The scattering time estimated from transport measurement is 9.6 $\times 10^{-10}$ s using $\mu_{tr}$ = $e\tau_{tr}$/$m^*$. There are nearly four order difference between the transport times ($\tau_{tr}$) and the quantum life time ($\tau_{xx}$), similar result has been observed in Cd$_3$As$_2$ and TaAs,\cite{LiangT, JiaS} one of the reason is that $\tau_{tr}$ measures backscattering processes that relax the current while $\tau_{xx}$ is sensitive to all processes that broaden the Landau levels.

Using the effective mass obtained above, we can calculate the electronic specific heat in quasi-two-dimensional approximation:\cite{Ca3Ru2O7}
\[\gamma_{N} = \sum_{i}\frac{\pi N_{A}k_{B}^{2}ab}{3\hbar^{2}}m^*\]
where $N_A$ is Avogadro's number, $k_B$ is Boltzmann's constant, $a$ and $b$ are the in-plane lattice parameters, $m^*$ is the quasiparticle mass and $\hbar$ is Planck's constant. From the effective mass obtained by quantum oscillation, and four  bands observed by ARPES,\cite{CavaRJ} $\gamma_{N}$ = 2.16 mJ mol$^{-1}$ K$^{-2}$ can be obtained, in excellent agreement with the $\gamma_N$ derived from specific heat [Fig. 2(b)], consistent with ARPES measurement\cite{CavaRJ} and indicating that four  bands detected by SdH alone contribute to the electronic specific heat.

One of the key evidence for the existence of Dirac Fermions is the non-trivial Berry's phase.\cite{Ando} Figure 3(f) presents the fan diagram of the Landau index. According to the Lifshitz-Onsager quantization rule, LL index $n$ is related to the cross section of FS $S_F$ by $S_F$($\hbar$/$eB$)  = 2$\pi$($n$ + $\gamma$). As shown in Fig. 3(f), the peak and valley positions of $\rho_{xx}$ fall on a straight line, the fit gives $\gamma$ = 0.21. $\gamma$ should be zero for conventional metals but ($\pm$ 1/2) for Dirac fermions due to the nonzero Berry's phase associated with their cyclotron motion. Berry phase deviates from the exact $\pi$ value has also been observed on NbP, Bi$_2$Se$_3$, and Bi$_2$Te$_2$Se.\cite{NbP} The deviation indicates significant spin-orbit coupling, in whose presence $\gamma=(1/2)+gm^{*}/4m$, where $g$ is the $g$-factor, $m^{*}$ is the cyclotron mass, and $m$ the electron mass.\cite{Mikitik1,Mikitik2,Falkovskii}The presence of parabolic bands at the Fermi surface that also might contribute to quantum oscillations is unlikely since their contribution would have been revealed in specific heat measurement. The peaks in $\Delta$$\rho_{xx}$ should be "phase-shifted" by 90$^{\rm o}$ from the peaks in $\Delta$$\rho_{xy}$. As presented by Fig. 3(f), after the peaks and valleys of $\Delta$$\rho_{xy}$ are shifted in $n$ by 1/4, the Landau fan phase diagram of $\Delta$$\rho_{xx}$ and $\Delta$$\rho_{xy}$ agree with each other quite well, confirming the non-zero Berry's phase obtained from $\Delta$$\rho_{xx}$.

The Bi square nets in SrMnBi$_2$, and CaMnBi$_2$ host Dirac states with quasi-2D Fermi surface. We perform the field dependence of longitudinal resistance up to 35 T at different angles to study the topological structure of YbMnBi$_2$. The geometry of the measurement setup is shown in the inset of Fig. 4(a). Figure 4(a) presents angle-dependent oscillation component after background subtraction. The oscillation peaks shift systematically with the angle increase. We perform FFT on the data in Fig. 4(a), and the results is shown in Fig. 4(b). Two peaks can be observed in the low angle data, we only take the main peak into consideration. The positions of main peak are summarized in Fig. 4(c); the frequency increases with the angle tilt from zero. The angle dependence of the FFT peaks can be roughly fitted assuming dominant contribution of quasi-2D conducting states ($F(0)$/cos$\theta$) at the Fermi surface. However, as shown in Fig. 4(c), the ellipsoid function $F(0)$/($\varepsilon^2$sin$^2\theta$ + cos$^2\theta$)$^{1/2}$, offers an alternative description. This suggests the interplay of quasi-2D and bulk 3D states in the topological structure of the FS. In addition, a dip between -23$^{\rm o}$ to -31$^{\rm o}$ also confirms non-trivial nature of the YbMnBi$_2$ Fermi surface.

The MR of solids only responds to the extremal cross section of the Fermi surface along the field direction. For a 2D fermi surface, the MR will only respond to the perpendicular component the of the magnetic field Bcos($\theta$). Angle-dependent MR of YbMnBi$_2$ single crystal at $B$ = 9 T and $T$ = 2 K is shown in Fig. 4(d). MR shows two fold symmetry; when magnetic field parallels to the $c$ axis of the single crystal ($\theta$ = 0), the MR is maximized and it gradually decreases with the field titling away from $c$ axis. MR is minimized when the field is applied in the $ab$ plane. The curve can be fitted by a function of  $\vert$cos($\theta$)$\vert$, indicating that quasi-2D fermi surface dominate magnetotransport in YbMnBi$_2$.

\begin{figure}[ht]
\centering
\includegraphics[width=0.5\textwidth]{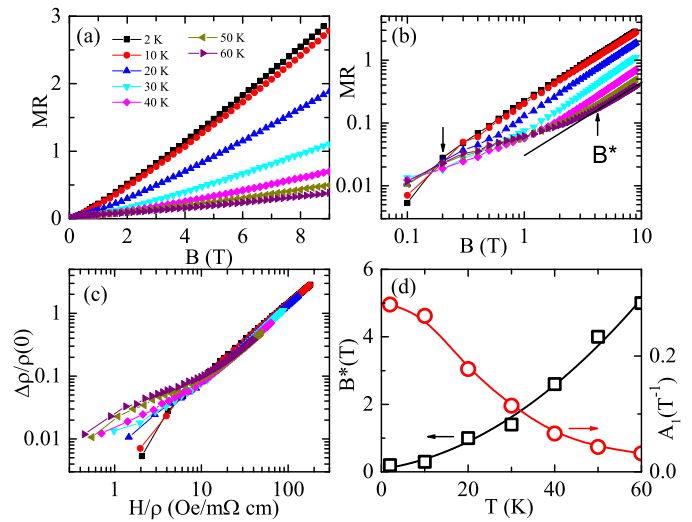}
\caption{(color online). (a) The magnetic field dependence of the in-plane MR at different temperatures. (b) Log-log plot of MR  versus magnetic field. (c) A Kohler plot for YbMnBi$_2$. (d) Temperature dependence of the critical field $B^*$ (black squares), the red solid line is the fitting results using $B^*$ = (1/2$e\hbar\upsilon_F^2$)($k_BT$+$E_F$)$^2$. Red circle corresponding to high-field MR linear coefficient $A_1$.} \label{fig4}
\end{figure}

Figure 5(a) presents the magnetic field dependence of MR at different temperatures. MR is also plotted in Fig. 5(b) on a log-log scale to emphasize the low field behavior. From Fig. 5(b), we can observe that linear MR behavior is established above a crossover field B$^*$. With increasing temperature, MR gradually decreases and $B^*$ increases. The normal MR of bands with parabolic dispersion either saturates at high fields or increases as $H^2$. The unusual nonsaturating linear magnetoresistance has been reported in Ag$_{2-\delta}$(Te/Se), Bi$_2$Te$_3$, Cd$_3$As$_2$, BaFe$_2$As$_2$ and (Ca, Sr)MnBi$_2$.\cite{KefengSr, KefengCa} In the quantum limit where all the carriers occupy only the lowest LL, the observed $B^*$ corresponds to the quantum limit of $B^*$ = (1/2$e\hbar\upsilon_F^2$)($k_BT$+$E_F$)$^2$.\cite{PRB 58 2788} As shown in Fig. 5(c), the B$^*$ can be fitted quite well by the above equation, which confirms the the existence of Dirac states in YbMnBi$_2$. Furthermore, MR in YbMnBi$_2$ conforms to Kohler's rule $\Delta$$\rho$/$\rho$(0) = $F$[$H$/$\rho$(0)] (where $F(H)$ usually follows a power law) in high magnetic fields [Fig. 5(c)]. This indicates that there is a single salient scattering time in YbMnBi$_2$, i.e. that even though the Fermi surface contains contribution from multiple bands, highly conducting (i.e. Dirac) states dominate MR.

\section{CONCLUSIONS}

 We have demonstrated quantum transport in antiferromagnetically ordered YbMnBi$_{2}$ single crystals. At 4 K the saturated magnetic moment is $4.3(1)\mu_B/$Mn whereas Mn magnetic moments are oriented along the c-axis. High-magnetic-field magnetotransport is consistent with the presence of Dirac fermions and significant spin-orbitcoupling of the Dirac-like carriers. It would be of interest to study in depth details of magnetic structure, Berry phase in ultrathin samples and putative ferromagnetic states in doped crystals.

\section*{Acknowledgements}

VOG gratefully acknowledge H. Cao for the support during the neutron diffraction experiment. Work at BNL was supported by the U.S. DOE-BES, Division of Materials Science and Engineering, under Contract No. DE-SC0012704. Work at the National High Magnetic Field Laboratory is supported by the NSF Cooperative Agreement No. DMR-0654118, and by the state of Florida. Work at the Oak Ridge National Laboratory was sponsored by the Scientific User Facilities Division, Office of Basic Energy Sciences, US Department of Energy. X-ray scattering data were collected at 28-ID-C x-ray powder diffraction beam line at National Synchrotron Light Source II at Brookhaven National Laboratory. Use of the National Synchrotron Light Source II, Brookhaven National Laboratory, was supported by the U.S. Department of Energy, Office of Science, Office of Basic Energy Sciences, under Contract No. DE-SC0012704.


\begin{references}

\bibitem{Dirac} T.O. Wehling, A.M. Black-Schaffer and A.V. Balatsky, Advances in Physics \textbf{63}, 1 (2014).
\bibitem{Vafek} Oskar Vafek and Ashvin Vishwanath, Annu. Rev. Condens. Matter Phys. \textbf{5}, 83 (2014).
\bibitem{PRL 104 137001} P. Richard, K. Nakayama, T. Sato, M. Neupane, Y.-M. Xu, J. H. Bowen, G. F. Chen, J. L. Luo, N. L. Wang, X. Dai, Z. Fang, H. Ding, and T. Takahashi, Phys. Rev. Lett. \textbf{104}, 137001 (2010).
\bibitem{Science 288 468} J. Orenstein and A. J. Millis, Science \textbf{288}, 468 (2000).
\bibitem{NovoselovK} K. S. Novoselov, A. K. Geim, S. V. Morozov, D. Jiang, M. I. Katsnelson, I. V. Grigorieva, S. V. Dubonos and A. A. Firsov, Nature \textbf{438}, 7065 (2005).
\bibitem{NM 6 183} A. K. Geim and K. S. Novoselov, Nat. Mater. \textbf{6}, 183 (2007).
\bibitem{RMP 81 109} A. H. Castro Neto, F. Guinea, N. M. R. Reres, K. S. Novoselov, and A. K. Geim, Rev. Mod. Phys. \textbf{81}, 109 (2009).
\bibitem{RMP 82 3045} M. Z. Hasan, and C. L. Kane, Rev. Mod. Phys. \textbf{82}, 3045 (2010).
\bibitem{RMP 83 1057} X. L. Qi and S. C. Zhang, Rev. Mod. Phys. \textbf{83}, 1057 (2011).
\bibitem{Fundamentals} A. A. Abrikosov, Fundamentals of the Theory of Metals (North-Holland, Amsterdam, 1988).
\bibitem{PRB 58 2788} A. A. Abrikosov, Phys. Rev. B \textbf{58}, 2788 (1998).
\bibitem{Nature 438 201} Y. Zhang, Z. Jiang, Y.-W. Tan, H. L. Stormer, and P. Kim, Nature (London) \textbf{438}, 201 (2005).
\bibitem{Science 324 924} D. Miller, K. Kubista, G. Rutter, M. Ruan,W. de Heer, P. First, and J. Stroscio, Science \textbf{324}, 924(2009).
\bibitem{PRL 106 217004} K. K. Huynh, Y. Tanabe, and K. Tanigaki, Phys. Rev. Lett. \textbf{106}, 217004 (2011).
\bibitem{KefengSr} Kefeng Wang, D. Graf, Hechang Lei, S. W. Tozer, and C. Petrovic, Phys. Rev. B \textbf{84}, 220401(R) (2011).
\bibitem{KefengCa} Kefeng Wang, D. Graf, Limin Wang, Hechang Lei, S. W. Tozer, and C. Petrovic, Phys. Rev. B \textbf{85}, 041101(R) (2012).
\bibitem{ParkSr} J. Park, G. Lee, F. Wolff-Fabris, Y. Y. Koh, M. J. Eom, Y. K. Kim, M. A. Farhan, Y. J. Jo, C. Kim, J. H. Shim, and J. S. Kim, Phys. Rev. Lett. \textbf{107}, 126402 (2011).
\bibitem{JiaLL} L.-L. Jia, Z.-H. Liu, Y.-P. Cai, T. Qian, X.-P. Wang, H. Miao, P. Richard, Y.-G. Zhao, Y. Li, D.-M. Wang, J.-B. He, M. Shi, G.-F. Chen, H. Ding and S.-C. Wang, Phys. Rev. B \textbf{90}, 035133 (2014).
\bibitem{PRB 87 245104} G. Lee, M. A. Farhan, J. S. Kim, and J. H. Shim, Phys. Rev. B \textbf{87}, 245104 (2013).
\bibitem{sr 4 5385} Y. Feng, Z. J. Wang, Ch. Y. Chen, Y. G. Shi, Z. J. Xie, H. M. Yi, A. J. Liang, S. L. He, J. F. He, Y. Y. Peng, X. Liu, Y. Liu, L. Zhao, G. D. Liu, X. L. Dong, J. Zhang, C. T. Chen, Z. Y. Xu, X. Dai, Z. Fang, and X. J. Zhou, Scientific Reports \textbf{4}, 5385 (2014).
\bibitem{Sales} A. F. May, M. A. McGuire, and B. C. Sales Phys. rev. B \textbf{90}, 075109 (2014).
\bibitem{Balents} Leon Balents, Physics \textbf{4}, 36 (2011).
\bibitem{CavaRJ} S. Borisenko, D. Evtushinsky, Q. Gibson, A. Yaresko, T. Kim, M. N. Ali, B. B\"{u}chner, M. Hoesch, R. J. Cava, arXiv: 1507.04847.
\bibitem{BurkovAA} A. A. Burkov, M. D. Hook, and Leon Balents, Phys. Rev. B \textbf{84}, 235126 (2011).
\bibitem{WanX} Xiangang Wan, Ari M. Turner, Ashvin Vishwanath, and Sergey Y. Savrasov, Phys. Rev. B \textbf{83}, 205101 (2011).
\bibitem{Hammersley} A. P. Hammersley, S. O. Swenson, M. Hanfland and D. Hauseman, High Pressure Res. \textbf{14}, 235 (1996).
\bibitem{Larson} A. C. Larson and R. B. von Dreele, Report No. LAUR-86-748, Los Alamos National Laboratory (1987).
\bibitem{Toby} B. H. Toby, J. Appl. Crystalogr. \textbf{34}, 210 (2001).
\bibitem{Gutman1} D. B. Gutman and D. L. Maslov, Phys. Rev. Lett. \textbf{99}, 196602 (2007).
\bibitem{Gutman2} D. B. Gutman and D. L. Maslov, Phys. Rev. B \textbf{77}, 035115 (2008).
\bibitem{GuoYF} Y. F. Guo, A.J. Princep, X. Zhang, P. Manuel, D. Khalyavin, I. I. Mazin, Y. G. Shi, and A. T. Boothroyd, Phys. Rev. B \textbf{90}, 075120 (2014).
\bibitem{Smith} R. A. Smith, \textit{Semiconductors} (Cambridge University, Cambridge, England, 1978).
\bibitem{NbP} C. Shekhar, A. K. Nayak, Y. Sun, M. Schmidt, m. Nicklas, I. Leermakers, U. Zeitler, Y. Skourski, J. Wosnitza, Z. K. Liu, Y. L. Chen, W. Schnelle, H. Borrmann, Y. Grin, C. Felser, and B. H. Yan, Nature Phys. \textbf{11}, 645 (2015).
\bibitem{LiangT} T. Liang, Q. Gibson, M. N. Ali, M. H. Liu, R. J. Cava, and N. P. Ong, Nature Mater. \textbf{14}, 280 (2015).
\bibitem{JoYJ} Y. J. Jo, J. Park, G. Lee, M. J. Eom, E. S. Choi, J. H. Shim, W. Kang, and J. S. Kim, Phys. Rev. Lett. \textbf{113}, 156602 (2014).
\bibitem{NC 5 5161} A. C. Potter, I. Kimchi, and A. Vishwanath, Nat. Commun. \textbf{5}, 5161 (2014).
\bibitem{Moll} P. J. W. Moll, N. L. Nair, T. Helm, A. C. Potter, I. Kimchi, A. Vishwanath, and J. G. Analytis, arXiv: 1505.02817
\bibitem{Shoeneberg} D. Shoeneberg, $Magnetic oscillation in Metals$ (Cambridge University Press, Cambridge, UK, 1984).
\bibitem{LiLJ} L. J. Li, K. F. Wang, D. Graf, L. M. Wang, A. F. Wang, and C. Petrovic, Phys. Rev. B \textbf{93} 115141 (2016).
\bibitem{JiaS} C. L. Zhang, Z. J. Yun, S. Y. Xu, Z. Q. Lin, B. B. Tong, M. Z. Hasan, J. F. Wang, C. Zhang, S. Jia, arXiv: 1502.00251
\bibitem{Ca3Ru2O7} N. Kikugawa, A. W. Rost, C. W. Hicks, A. J. Schofield, and A. P. Mackenzie, J. Phys. Soc. Jpn., \textbf{79}, 024704 (2010).
\bibitem{Ando} Yoichi Ando, J. Phys. Soc. Jpn. \textbf{82}, 102001 (2013).
\bibitem{Mikitik1} G. P. Mikitik and Yu. V. Sharlai, Phys. Rev. B \textbf{67}, 115114 (2003).
\bibitem{Mikitik2} G. P. Mikitik and Yu. V. Sharlai, Phys. Rev. B \textbf{85}, 033301 (2012).
\bibitem{Falkovskii} L. A. Falkovskii, J. Exp. Theor. Phys. \textbf{17}, 1302 (1963).













\end{references}
\end{document}